\documentstyle[preprint,aps,prl,epsf]{revtex}
\begin{document}
\preprint
\widetext
\title{Vertex-corrected tunneling inversion in superconductors: Pb
}
\author{
J. K. Freericks$^{(a)}$, E. J. Nicol$^{(b)}$, A. Y. Liu$^{(a)}$, and
A. A. Quong$^{(c)}$.
}
\address{
$^{(a)}$Department of Physics,
Georgetown University, Washington, DC 20057-0995\\
$^{(b)}$ Department of Physics,
University of Guelph, Guelph, ON N1G 2W1, Canada\\
$^{(c)}$ Sandia National Laboratories, Livermore, CA 94551-0969 }
\date{\today}
\maketitle
\begin{abstract}
The McMillan-Rowell tunneling inversion program, which extracts the 
electron-phonon spectral function $\alpha^2F(\Omega)$ and the Coulomb
pseudopotential $\mu^*$ from experimental tunneling data, is generalized
to include the lowest-order vertex correction.  We neglect 
the momentum dependence of the electron-phonon matrix elements,
which is equivalent to using a local approximation. The perturbation theory 
is performed on the imaginary axis and then an exact analytic continuation 
is employed to produce the density of states on the real axis.  Comparison 
is made with the experimental data for Pb.
\end{abstract}
\renewcommand{\thefootnote}{\copyright}
\footnotetext{ 1996 by the authors.  Reproduction of this article by any means
is permitted for non-commercial purposes.}
\renewcommand{\thefootnote}{\alpha{footnote}}
\pacs{74.20.-z, 71.27.+a, 71.38.+i}

\narrowtext
The theory of low-temperature superconductors is one of the most accurate
theories in condensed-matter physics.  Agreement to better than
one part in $10^4$ is common between the tunneling density of states (DOS)
measured experimentally and that calculated with an extracted electron-phonon
spectral function $\alpha^2F(\Omega)$ and Coulomb pseudopotential $\mu^*$.
The reason why the agreement is so good is due to Migdal's theorem 
\cite{migdal},
as formulated by Eliashberg \cite{eliashberg} in the superconducting state,
which says that there is a small parameter in the theory, namely the ratio
of the phonon energy scale to the electronic energy scale, that guarantees
the rapid convergence of the perturbative
expansion.  However in some recently discovered materials,
that are believed to be electron-phonon superconductors, the ratio of the phonon
to electronic energy scale is no longer as small.  
Two examples are Ba$_{1-x}$K$_x$BiO$_3$
\cite{bkboexp} and the doped fullerenes \cite{c60exp} which have relatively high
transition temperatures.
This motivates the need for a theory that includes the effects
of the so-called vertex corrections, which are neglected in the 
conventional Migdal-Eliashberg
formalism.  In this contribution, we present a generalization of 
the McMillan-Rowell \cite{mcmillan_rowell} tunneling-inversion program
to include the lowest-order effects of vertex corrections in the local
approximation, and apply the new formalism to the low-temperature superconductor
Pb to illustrate how vertex corrections can be incorporated into such an
analysis, and to determine, quantitatively, the accuracy of the 
Migdal-Eliashberg formalism.

Recent work on incorporating vertex corrections into the theory of 
superconductivity \cite{sham,kostur,pietronero,freericks_vert} has illustrated
some of their qualitative effects:  If one adopts
the conventional approximation of assuming a constant electronic DOS and
neglecting the momentum dependence of the electron-phonon matrix elements,
then the vertex corrections will suppress $T_c$ and the isotope coefficient
$\alpha$.  Incorporation of either momentum dependence to the 
matrix elements, or nonconstant DOS can lead to enhancements to $T_c$ and 
$\alpha$.  Little is known about how large these effects can be in real
materials, but they have been verified for model systems by comparing 
vertex-corrected theories \cite{freericks_vert}
to the exact solution of electron-phonon models in 
the infinite-dimensional limit \cite{freericks_qmc}.

Even a well-studied low-temperature superconductor such as Pb has a
nagging inconsistency between the experimentally extracted $\alpha^2F$
and $\mu^*$ and the bulk transition temperature $T_c$.  
The extracted dimensionless electron-phonon coupling strength $\lambda$
satisfies $\lambda=1.55$, and the Coulomb pseudopotential is $\mu^*=0.131$
\cite{mcmillan_rowell}.  However, the Coulomb pseudopotential must be
increased to $\mu^*=0.144$ \cite{reviews} in order to produce the correct
$T_c$ of 7.19K.  It is possible that including the vertex corrections can 
explain this discrepancy.

Furthermore, new computational strategies have been developed that greatly
improve the accuracy of the Migdal-Eliashberg formalism, and allow for
a straightforward generalization to the incorporation of the lowest-order
vertex corrections: (i) use of the local approximation for the many-body
problem \cite{metzner_vollhardt}; (ii) performing exact analytic continuations
from the imaginary axis to the real axis \cite{marsiglio_ac}; and (iii)
incorporation of high-frequency resummation schemes \cite{marsiglio_hf,deisz}.
We employ all three strategies in our computational methods.

The electronic self-energy and the irreducible vertex function for
superconducting order including the lowest-order vertex correction beyond
the conventional model are both illustrated in Figure~1.  The wavy lines
denote dressed phonon propagators, and the solid lines denote dressed
electronic Green functions in the Nambu-Gorkov formalism.  We make the
conventional approximations \cite{reviews}: (i) neglect the angular
dependence of the electron-phonon matrix elements and the phonon spectral
functions, and evaluate them at the Fermi energy; (ii) neglect the energy
dependence of the electronic density of states $\rho(\epsilon)$ and 
evaluate it at the Fermi energy $\rho(0)$; and (iii) treat the Coulomb 
interactions via a pseudopotential for the anomalous self-energy.  
When these approximations are invoked, one must solve a self-consistent
perturbation theory in frequency-space only---the self-energy is replaced
by a momentum-independent function that has been averaged over the Fermi
surface.  {\it These self-consistent equations are identical in form to the
equations one would derive in the local approximation, valid in the 
large-dimensional limit \cite{metzner_vollhardt,freericks_vert}.  }

The perturbation theory is performed on the imaginary axis at the electronic
Matsubara frequencies $\omega_n:=\pi T(2n+1)$.  This allows for a
proper treatment of the Coulomb pseudopotential, since the sharp cutoff
lies on the imaginary, not the real axis \cite{fenton}. If we make the 
conventional definitions for the quasiparticle renormalization
$Z_n:=Z(i\omega_n)=1-{\rm Im}[\Sigma(i\omega_n)]/\omega_n$ and for
the superconducting gap $\Delta_n:=\Delta(i\omega_n)=\phi(i\omega_n)/Z(i
\omega_n)$, then the self-consistent equations are:
\begin{eqnarray}
Z_n&=&1+\frac{\pi T}{\omega_n}\sum_{l=-N}^N\lambda_l\frac{\omega_{n-l}}
{\sqrt{\omega_{n-l}^2+\Delta_{n-l}^2}}+\delta Z_n\cr
&+&\frac{\pi^3T^2C\rho(0)}{\omega_n}\sum_{l=-N}^N\sum_{l^{\prime}=-N}^N
\lambda_l\lambda_{l^{\prime}}\frac{-\omega_{n-l}\omega_{n-l^{\prime}}
\omega_{n-l-l^{\prime}}-2\omega_{n-l}\Delta_{n-l^{\prime}}
\Delta_{n-l-l^{\prime}}+\omega_{n-l-l^{\prime}}\Delta_{n-l}
\Delta_{n-l^{\prime}}}
{\sqrt{(\omega_{n-l}^2+\Delta_{n-l}^2)(\omega_{n-l^{\prime}}^2+\Delta_{n-
l^{\prime}}^2)(\omega_{n-l-l^{\prime}}^2+\Delta_{n-l-l^{\prime}}^2)}}\quad ,
\label{eq: zn}
\end{eqnarray}
\begin{eqnarray}
\Delta_nZ_n&=&\pi T\sum_{l=-N}^N(\lambda_l-\mu^*)\frac{\Delta_{n-l}}
{\sqrt{\omega_{n-l}^2+\Delta_{n-l}^2}}\cr
&+&\pi^3T^2C\rho(0)\sum_{l=-N}^N\sum_{l^{\prime}=-N}^N
\lambda_l\lambda_{l^{\prime}}\frac{-\Delta_{n-l}\Delta_{n-l^{\prime}}
\Delta_{n-l-l^{\prime}}-2\Delta_{n-l}\omega_{n-l^{\prime}}
\omega_{n-l-l^{\prime}}+\omega_{n-l}\omega_{n-l^{\prime}}\Delta_{n-l-l^{\prime}}
}{\sqrt{(\omega_{n-l}^2+\Delta_{n-l}^2)(\omega_{n-l^{\prime}}^2+\Delta_{n-
l^{\prime}}^2)(\omega_{n-l-l^{\prime}}^2+\Delta_{n-l-l^{\prime}}^2)}}\quad ,
\label{eq: deltan}
\end{eqnarray}
where $\lambda_n:=2\int_0^{\infty} d\Omega\alpha^2F(\Omega)\Omega/(\Omega^2+
4\pi^2T^2n^2)$ is the dimensionless electron-phonon coupling, $N=\frac{1}{2}
(\frac{\omega_c}{\pi T}-1)$ is the cutoff for the summations [the frequency
cutoff is chosen to be six times the maximal frequency in $\alpha^2F(\Omega)$, 
or $\omega_c=6\omega_{max}$], $\delta Z_n$ is defined below, and $C$ is a 
Fermi-surface average for the vertex-correction terms.  This average is 
defined by
\begin{equation}
C:=\frac{1}{\rho^4(0)}\sum_q\left [ \sum_k\delta(\epsilon_F-\epsilon_k)
\delta(\epsilon_F-\epsilon_{q-k})\right ]^2
\label{eq: cdef}
\end{equation}
with $\epsilon_F$ the Fermi energy and $\epsilon_k$ the band structure.
In a free-electron model, with a $k^2$ dispersion, the constant $C$
assumes the form $C=\frac{1}{6n}$, with $n$ the number of free electrons per
spin per unit cell \cite{kostur}. Since the DOS at the Fermi level for a
free-electron model is $\rho(0)=\frac{3n}{2\epsilon_F}$, the product $C\rho(0)$
is $\frac{1}{4\epsilon_F}$ for a free-electron model.  For Pb, we perform 
a scalar relativistic density-functional calculation of the band structure in 
the local-density approximation, and find $\rho(0)=2.5\times 10^{-4}$ 
states/spin/meV, and $C=0.15$, yielding $C\rho(0)=4\times 10^{-5}$ 
states/spin/meV.  This is the value we use for the numerical work.

Finally, a high-frequency resummation scheme \cite{marsiglio_hf,deisz} is 
employed,
that calculates the perturbation-theory results relative to the exact results
for the normal state, namely
\begin{equation}
Z_n({\rm normal})=1+\frac{1}{2n+1}[\lambda_0+2\sum_{m=1}^n\lambda_m]\quad ,
\label{eq: znnorm}
\end{equation}
by appending the normal state results to the perturbation theory illustrated in 
Figure~1.  This is achieved by adding the contribution
\begin{equation}
\delta Z_n:=Z_n({\rm normal})-1-\frac{1}{2n+1}\sum_{m=-N}^{N}\lambda_m
{\rm sgn}(n-m-\frac{1}{2})\quad ,
\end{equation}
[where ${\rm sgn}(x)$ is 1 if $x>0$ and $-1$ if $x<0$] to the 
perturbation series.  The term $\delta Z_n$ is the normal-state contribution
to the quasiparticle renormalization factor that is usually neglected when
one introduces the cutoff $N$ into the frequency summations.

The difficult step in the tunneling-inversion program is to analytically
continue the gap function to the real axis, in order to determine the
tunneling DOS $N(\omega)$ from
\begin{equation}
N(\omega)=N(0){\rm Re}\left [\frac{\omega}{\sqrt{\omega^2-\Delta^2(\omega)}}
\right ]\quad .
\label{eq: tundos}
\end{equation}
Recently, however, it was discovered that an exact analytic continuation could
be performed by following the prescription of Baym and Mermin
\cite{baym}: formally
perform the analytic continuation $i\omega_n\rightarrow \omega+i\delta$,
and then add a function that vanishes at each of the Matsubara frequencies,
and is bounded in the upper half-plane except for simple poles located
at just the positions necessary to cancel the poles introduced by the
formal analytic continuation.  Since the analytic continuation is unique,
and since the final function can be shown to be analytic in the upper
half-plane, this analytic-continuation procedure is exact.  Such a scheme
has already been implemented for the Migdal-Eliashberg theory 
\cite{marsiglio_ac}, and it is a straightforward but tedious procedure
to generalize these results
to include the lowest-order vertex corrections.  The end result
is a formula for the quasiparticle renormalization, and the gap, on the real
axis, that involves the data on the imaginary axis and integrals of
the Green function evaluated in the upper half-plane.  Because of the dependence
on the Green function, these equations require a self-consistent solution
on a real-axis grid (a step size of 0.05~meV is chosen for the grid spacing).
The final equations are cumbersome and will be shown elsewhere.  

Finally, we use this formalism to extract both $\alpha^2F(\Omega)$
and $\mu^*$ from the experimental data.  We follow the original prescription
of McMillan and Rowell \cite{mcmillan_rowell}: (i) Guess an initial value
for $\alpha^2F(\Omega)$. (ii) Adjust $\mu^*$ to reproduce the experimental
superconducting gap at zero temperature $\Delta_0$ [which is defined from
${\rm Re} \Delta(\omega)=\omega$ at $\omega=\Delta_0$]. An analytic continuation
employing a Pad\' e approximation is used to determine $\mu^*$, since 
$\Delta_0$ is determined to an accuracy of one part in $10^5$ with such an
approximation.  (iii) Compute the functional derivative of the change in the 
tunneling DOS with respect to a change in $\alpha^2F(\Omega)$. (iv)
Determine the shift in $\alpha^2F$ by solving the matrix equation
\begin{equation}
\int d\Omega\frac{\delta N(\omega)}{\delta \alpha^2F(\Omega)}\delta 
\alpha^2F(\Omega)=N(\omega)-N_{exp}(\omega)\quad .
\label{eq: funder}
\end{equation}
This expression is discretized on the real axis, and a 
singular-value-decomposition is employed to determine the shift
$\delta\alpha^2F(\Omega)$,
since there are small eigenvalues of the functional derivative matrix,
which cause instabilities in updating $\alpha^2F$.  (v) Determine the new 
$\alpha^2F(\Omega)$ by adding a smoothed shift $\delta\alpha^2F(\Omega)$
to it, with $\alpha^2F$ forced to behave quadratically in $\Omega$ for
$\Omega<0.5$~meV.  This procedure is iterated until convergence is reached.

The results for the tunneling inversion for Pb for both the Migdal-Eliashberg
theory, and the vertex-corrected theory are presented in Table~1.  Various 
parameters are recorded including $\lambda$, the ``average'' phonon frequency
$\omega_{ln}:=\exp[2\int_0^{\infty}d\Omega\ln (\Omega)\alpha^2F(\Omega)/
\Omega\lambda]$, the area $A$ under $\alpha^2F$, $\mu^*$, $\Delta_0$,
$T_c$, the maximum error, and the root-mean-square error of the fit.
Note how a proper treatment of the cutoff for $\mu^*$ and inclusion of the
high-frequency resummation improves the calculated $T_c$ and that the
vertex corrections modify $\lambda$ by the order of 1\% even though the
Migdal parameter [$C\rho(0)\omega_{max}\lambda$] is on the order of 0.0007. 
A plot of the extracted $\alpha^2F(\Omega)$ is
given in Figure~2(a).  The vertex-corrected fit is the solid line and the
Migdal-Eliashberg fit is the dashed line.  The difference between the two
spectral functions $\alpha^2F_V(\Omega)-\alpha^2F_{ME}(\Omega)$ is plotted
in Figure~2(b). The vertex corrections produce
slight enhancements to the transverse and longitudinal phonon peaks and they
suppress the spectral weight in the region beyond the maximal bulk phonon
frequency for Pb (about 9 meV). Finally a comparison of the
fitted tunneling DOS is compared to the experimental data 
\cite{mcmillan_rowell} in Figure~3.
The small dots are the experimental data points used in the fitting procedure;
the larger dots (from data taken at a higher temperature) were not used
in the fit.  The solid line is the vertex-corrected DOS and the dotted line
is the Migdal-Eliashberg DOS.  Note that the two curves lie on top of
each other for low energy, but deviate more at higher energies.  It is
difficult to tell which fit is more accurate.

We conclude with a discussion of the general properties of vertex corrections.
In the conventional model, where the momentum dependence of the electron-phonon
matrix elements is neglected, and the electronic DOS is assumed to be constant,
the vertex corrections will always cause a reduction to $T_c$ and the 
isotope coefficient $\alpha$.  A generalization of the formula for the Holstein
model \cite{freericks_vert} shows that the ratio of the vertex-corrected
$T_c$ to the Migdal-Eliashberg $T_c$ is
\begin{equation}
\frac{T_c({\rm vertex})}{T_c({\rm no~vertex})}=
\exp\left [-\frac{2\pi^2C\rho(0)}{\lambda}\left (
\int_0^{\infty}d\Omega\alpha^2F(\Omega)+\frac{1}{\lambda}
\int_0^{\infty}d\Omega\alpha^2F(\Omega)\int_0^{\infty}d\Omega^{\prime}
\alpha^2F(\Omega^{\prime})\frac{1}{\Omega+\Omega^{\prime}}\right )\right ],
\label{eq: tc_red}
\end{equation}
in the weak-coupling limit (if we set $\mu^*=0$).  This formula gives an 
order-of-magnitude
estimate for when effects of vertex corrections should be important
in a real material (for Pb the ratio is 0.9978).

The other main effect of vertex corrections is that they will modify the
higher-energy structure in $\alpha^2F(\Omega)$, because they involve processes
where two phonons scatter, so that the structures in $\alpha^2F(\Omega)$
are changed at multiples of the lower-energy peaks.  
This effect is small in a low-temperature superconductor, but could be
significant in the newly discovered superconductors.

Finally, the vertex corrections will also modify the isotope coefficient.
It is possible that effects of vertex corrections could be seen upon
reexamination of isotope coefficient data on low-temperature superconductors.
Work in this direction is currently in progress.  We also plan on extracting
$\alpha^2F$ and $\mu^*$ for other low-temperature superconductors to see
if the effects of vertex corrections are important in other materials.  We
plan on applying the vertex-corrected tunneling inversion program to
both Ba$_{1-x}$K$_x$BiO$_3$ and the doped fullerenes once accurate
tunneling data for these materials becomes available.

We would like to acknowledge useful discussions with 
J. Carbotte,
D. Hess,
M. Jarrell,
V. Kostur,
F. Marsiglio,
B. Mitrovi\' c,
J. Rowell,
D.\ Scalapino, 
H.-B. Sch\" uttler,
and J. Serene.
J.~K.~F. acknowledges the
Donors of The Petroleum Research Fund, administered by the
American Chemical Society 
(ACS-PRF No. 29623-GB6) and the Office of Naval Research Young Investigator
Program (N000149610828) for partial support of this research.
E.~J.~N acknowledges support from the Natural Sciences and Engineering
Research Council of Canada (NSERC).

\mediumtext
\begin{table}
\caption{
Comparison of fitted tunneling inversion data for the Migdal-Eliashberg theory
and the vertex-corrected theory. The vertex corrections modify $\lambda$ by
the order of 1\%.}
\begin{tabular} {rcccccccc}
Theory &$\lambda$&$\omega_{ln}$ (meV)&$A$ (meV)
&$\mu^*$&$\Delta_0$ (meV)&$T_c (K)$&max. error&r.m.s. error\\
\tableline
Migdal-Eliashberg&1.542&4.863&4.029&0.136&1.400&7.23&0.0004&0.0001\\
Vertex-Corrected&1.561&4.847&4.070&0.141&1.400&7.22&0.0007&0.0001\\
\end{tabular}
\end{table}

\begin{figure}[t]
\epsffile{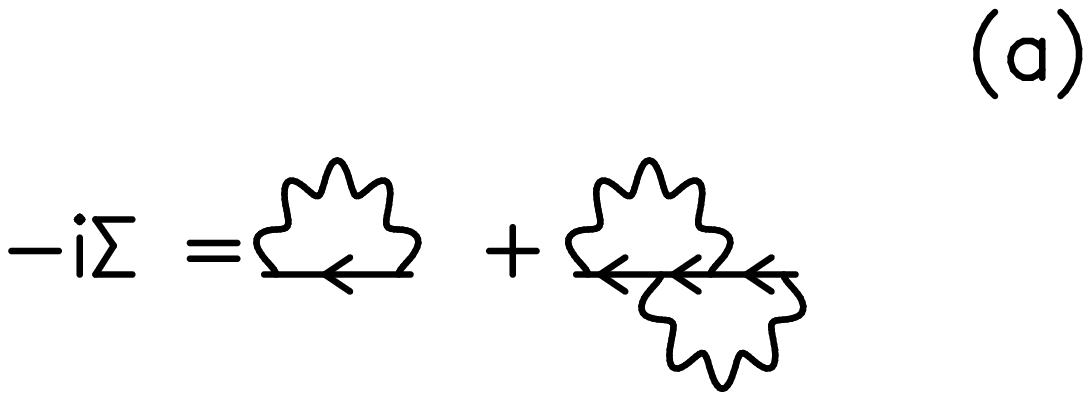}
\epsffile{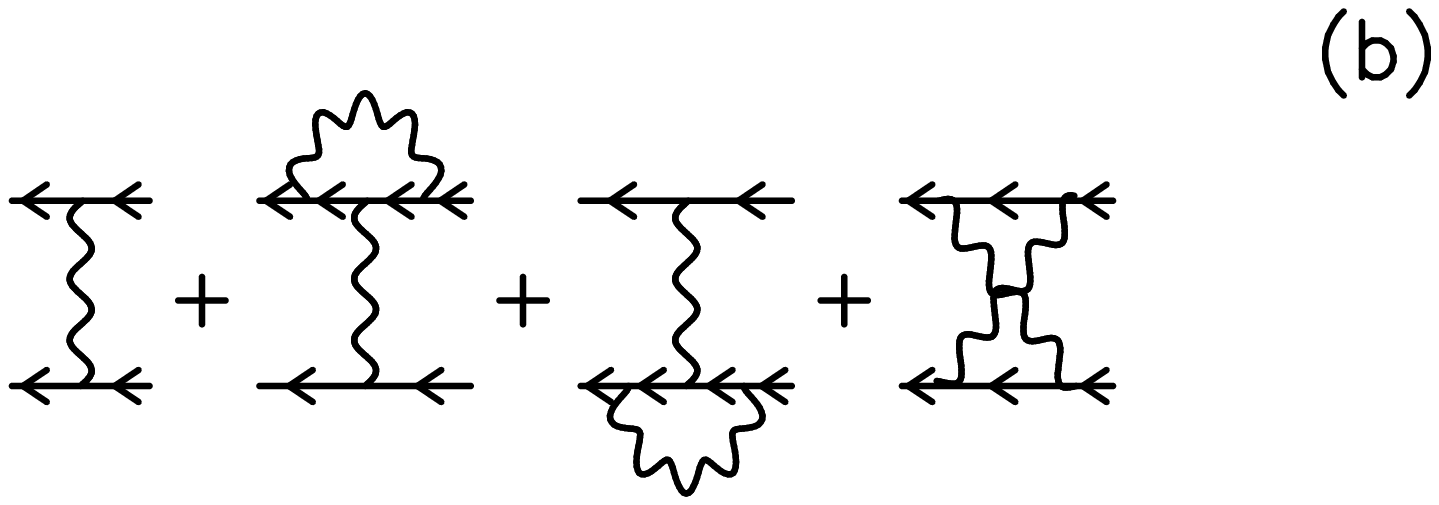}
\caption{Feynman diagrams for (a) the electronic self-energy and (b)
the irreducible vertex function in the superconducting channel.
The first diagram in (a) and (b) is the Migdal-Eliashberg approximation, the 
remaining diagrams are the lowest-order vertex corrections.
The solid lines denote dressed electron propagators and the wiggly lines
denote dressed phonon propagators. }
\end{figure}

\pagebreak

\begin{figure}[t]
\epsfxsize=5.0in
\epsffile{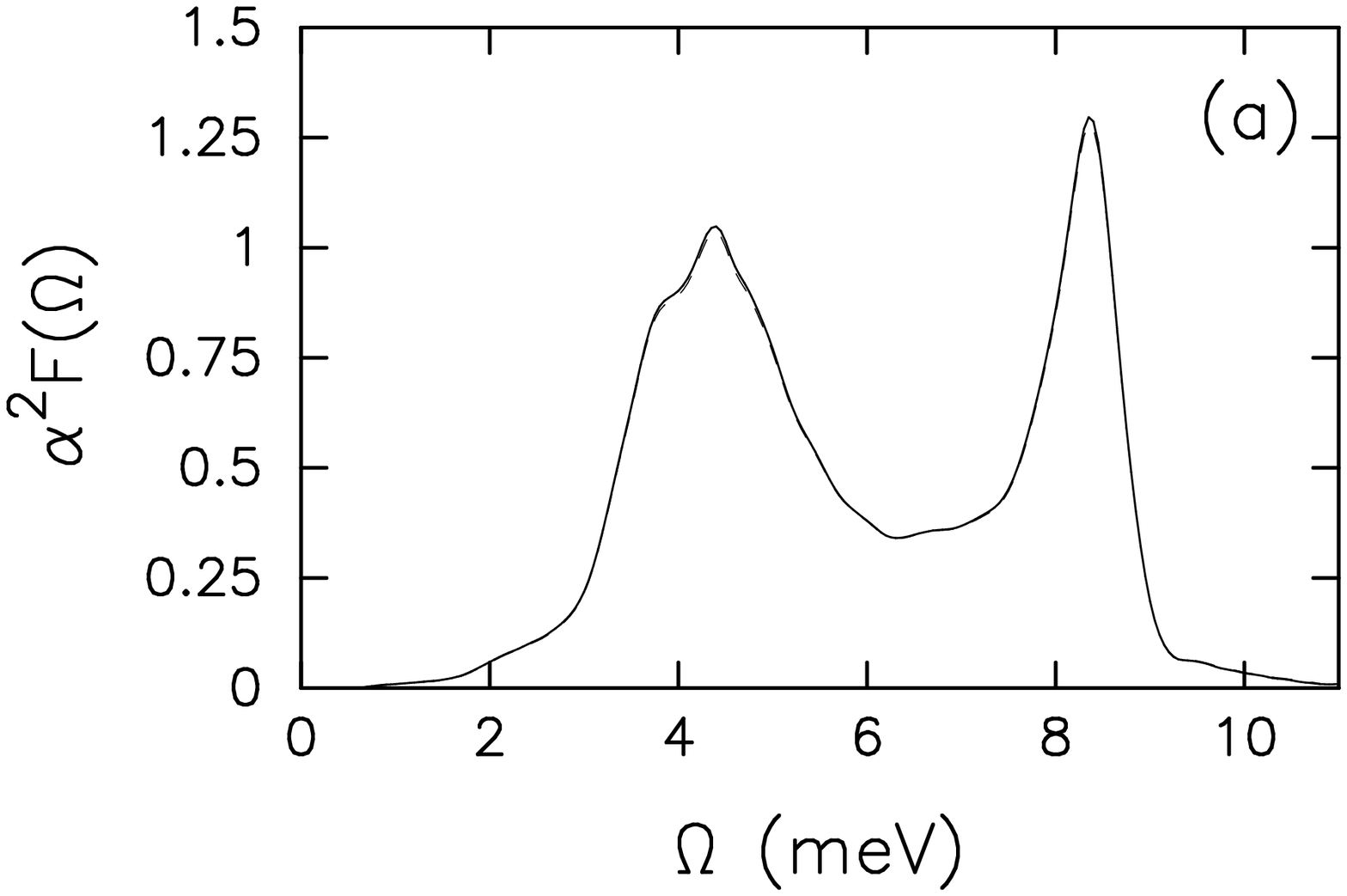}
\epsfxsize=5.0in
\epsffile{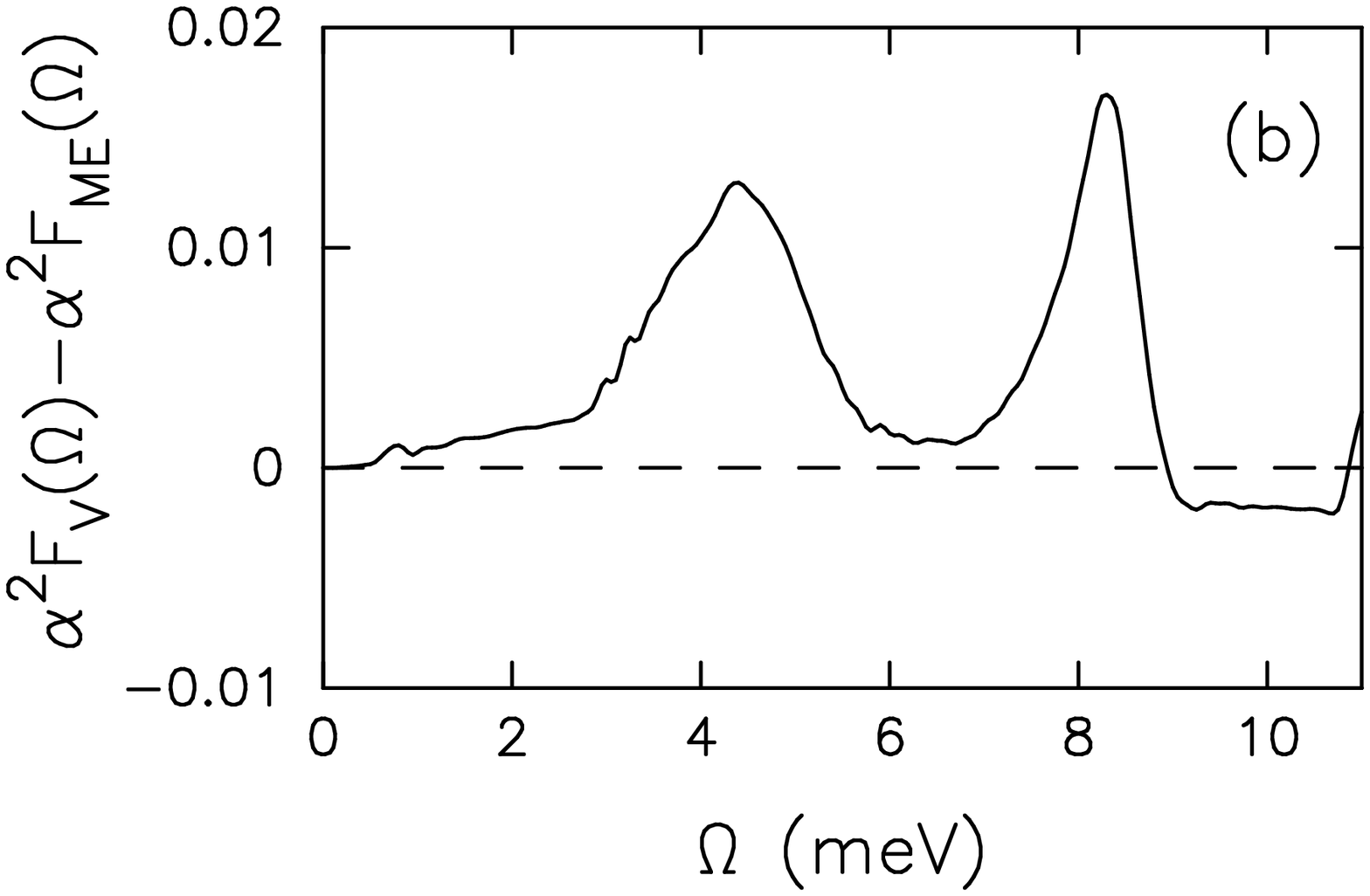}
\caption{(a) 
Electron-phonon spectral function, $\alpha^2F(\Omega)$, extracted from
the experimental tunneling data for Pb.  The solid line is the vertex-corrected
fit and the dashed line is the Migdal-Eliashberg fit. (b) Difference in 
extracted spectral functions $\alpha^2F_V(\Omega)-\alpha^2F_{ME}(\Omega)$.
Note the enhancement of the peaks and the suppression in the region where
$\Omega$ is larger than the maximum phonon frequency in bulk Pb.  }
\end{figure}

\pagebreak

\begin{figure}[t]
\epsfxsize=5.0in
\epsffile{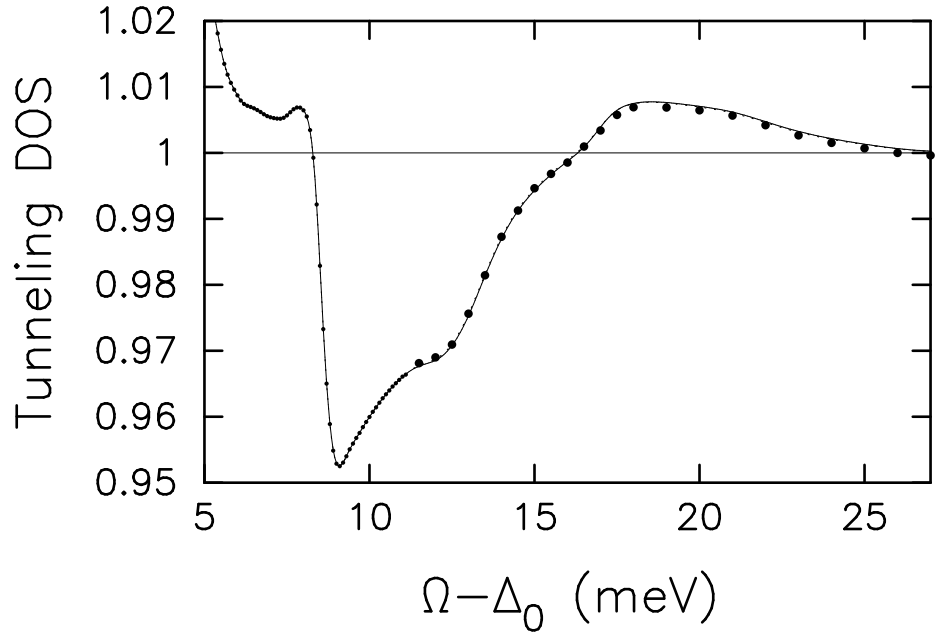}
\caption{Comparison of the experimental tunneling DOS (solid dots) to the
fitted DOS for Pb.  The solid line is the vertex-corrected 
fit and the dotted line is the Migdal-Eliashberg fit. The small dots are the
experimental data included in the fit, while the large dots are experimental
data not included in the fit, but taken at a higher temperature (and hence less
accurate). }
\end{figure}

\end{document}